\input{aipcheck}

\documentclass[
    ,final            % use final for the camera ready runs
  ]
  {aipproc}

\layoutstyle{6x9}

\begin{document}

\title{SUSY-QCD corrections to dark matter annihilations}


\author{K.~Kova\v{r}\'{\i}k}{
  address={Laboratoire de Physique Subatomique et de Cosmologie,\\ Universit\'{e} Joseph Fourier Grenoble, CNRS/IN2P3, Institut Polytechnique de Grenoble\\ 53 rue des Martyrs, 38026 Grenoble, France}
}

\author{B.~Herrmann}{
  address={Institut f\"ur Theoretische Physik und Astrophysik,
 Universit\"at W\"urzburg,
 Am Hubland, D-97074 W\"urzburg, Germany}
}

\begin{abstract}
	We present full QCD and SUSY-QCD corrections to neutralino pair annihilation
	into quark-antiquark pairs. We show results and numerically
	evaluate their impact on the neutralino annihilation cross section and the relic density in scenarios 
	with non-universal gaugino masses. 
	By comparing to the relic density obtained from a pure leading order
	calculation, we demonstrate that the corrections strongly influence the
	extraction of SUSY mass parameters from cosmological data.\end{abstract}
\maketitle

%%%%%%%%%%%%%%%%%%%%%%%%%%%%%%%%%%%%%%%%%%%%
%% MAINMATTER
%%%%%%%%%%%%%%%%%%%%%%%%%%%%%%%%%%%%%%%%%%%%

\section{Introduction}
The cosmological observations such as the mission of the Wilkinson Microwave
Anisotropy Probe (WMAP) which indicate the existence of Cold Dark Matter (CDM) 
in the universe, provide first hints of physics beyond the Standard Model (SM) 
as all SM particles are excluded to be the dark matter candidate. On top of that the 
measurement of the relic density of dark matter has become so precise that it turned into 
a very powerful tool to test models beyond the Standard Model which contain a suitable 
candidate for dark matter. A combination of the five-year measurement of the
cosmic microwave background by the WMAP mission with supernova and baryonic
acoustic oscillation data yields the narrow $2\sigma$-interval for the relic
density of dark matter \cite{Hinshaw:2008kr}
\begin{equation}\label{cWMAP}
  0.1097 < \Omega_{\rm CDM}h^2 < 0.1165\,,
\end{equation}
where $h$ denotes the present Hubble expansion rate $H_0$ in units of 100 km
s$^{-1}$ Mpc$^{-1}$.

The relic density of the dark matter particle $\chi$ is 
directly proportional to its present number density $n_0$. It is obtained by solving the Boltzmann equation
describing the time evolution of the number density
\begin{equation}
 \frac{dn_{\tilde{\chi}}}{dt} ~=~ -3 H n_{\tilde{\chi}} - \langle
 \sigma_{\rm ann}v \rangle \big( n_{\tilde{\chi}}^2 - n_{\rm eq}^2 \big) \,,
\end{equation}
where the first term on the right-hand side corresponds to a dilution due to the
expansion of the universe, and the second term corresponds to a decrease due to
annihilations and co-annihilations of the relic particle into SM particles
\cite{Gondolo:1990dk}. The Boltzmann equation can be integrated numerically and the energy density of the dark matter particle is approximately inversely proportional to the thermally averaged annihilation cross-section $\langle \sigma_{\rm ann}v \rangle$.

Here, we use the precise experimental measurement of dark matter's relic density to constrain parameter 
space of the Minimal Supersymmetric Standard Model (MSSM).  In order to keep up with current and future
experimental improvements, one has to reduce the theoretical uncertainty involved in the analysis, which 
we do by including the higher-order supersymmetric QCD corrections to annihilation processes 
with quark final states (for details see Ref.~\cite{Herrmann:2009mp}).

In many scenarios of the MSSM, the dark matter candidate is the lightest neutralino.
In our analysis, we concentrate on scenarios that are derived from minimal supergravity (mSUGRA) models, where 
heavy quark final states dominate the annihilation. We choose to relax the gaugino mass unification 
which allows to investigate scenarios where the cross-section is dominated by the 
exchanges of the $Z^0$-boson or the $t$-channel scalar quark exchanges as opposed to the Higgs-boson exchange in mSUGRA (see Refs.~\cite{Bertin:2002sq,Martin:2007gf,Baer:2007uz}).

The soft SUSY-breaking Lagrangian containing gaugino mass terms for the $B$-ino, $W$-ino, and gluino has the following form
\begin{equation}
  {\cal L}_{\rm soft} ~\subset~ -\frac{1}{2} \Big( M_1 \tilde{B}\tilde{B} + M_2 \tilde{W}\tilde{W} + M_3 \tilde{g}\tilde{g} + {\rm h.c.} \Big) .
\end{equation}
Relaxing the universality of gaugino masses at the GUT scale, the values of $M_1$, $M_2$, and $M_3$ at the unification scale can be considered as independent parameters.  The sfermion and Higgs sectors are parametrized by the usual mSUGRA parameters $m_0$, $m_{1/2}$, $A_0$, $\tan\beta$, and $\rm{sgn}(\mu)$ at the unification scale.
Here, we adopt a commonly used parametrization and introduce the dimensionless parameters
\begin{equation}
  x_1 ~=~ \frac{M_1}{M_2} \qquad {\rm and}\qquad x_3 ~=~ \frac{M_3}{M_2},
\end{equation}
which will be used together with the $W$-ino mass parameter $M_2$ to describe the gaugino sector. The case $x_1=x_3=1$ reproduces the mSUGRA model.
\begin{table}
  \begin{tabular}{c|ccccc|cc|cc|}
    & $m_0$ [GeV] & $M_2$ [GeV] & $A_0$ [GeV] & $\tan\beta$ & $\rm{sgn}(\mu)$ & $x_1$ & $x_3$ & $\Omega_{\rm CDM}h^2$ & $t\bar{t}$  \\
    \hline
    I & 1500 & 600 & 0 & 10 & + & 1 & 4/9 & 0.104 & 50.4\%\\
    II & 320 & 700 & -350 & 10 & + & 2/3 & 1/3 & 0.114 & 79.2\%\\
    \hline
  \end{tabular}
\caption{High-scale parameters together with the corresponding neutralino relic density, the contribution from top quark-antiquark final states to the annihilation cross section obtained with {\tt micrOMEGAs 2.1} \cite{Belanger:2008sj} for our two selected non-universal gaugino mass scenarios.}
\label{tab:a}
\end{table}
\section{Results}
\begin{figure}
\begin{tabular}{cc}
    \includegraphics[scale=0.3]{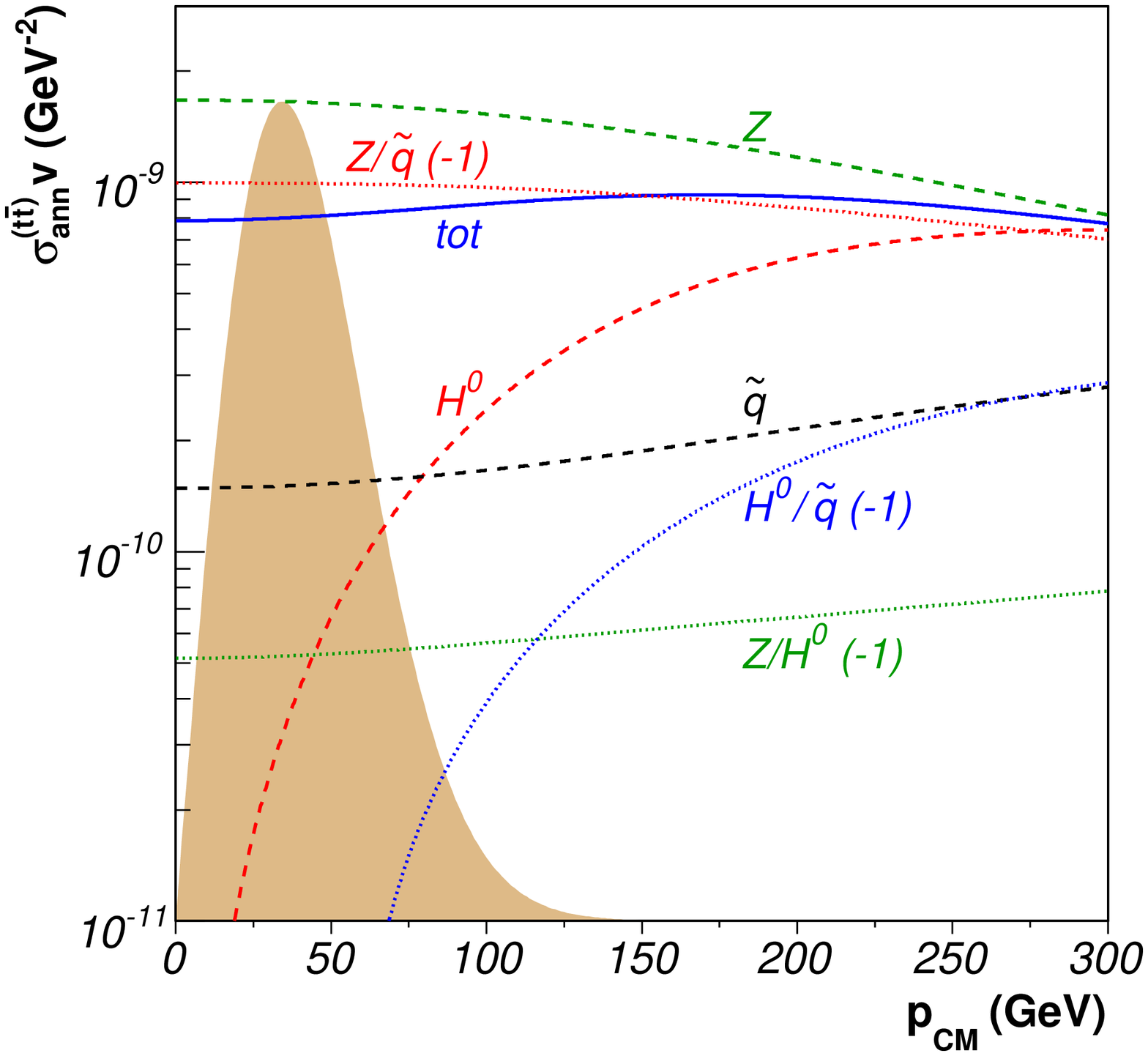} &
    \includegraphics[scale=0.3]{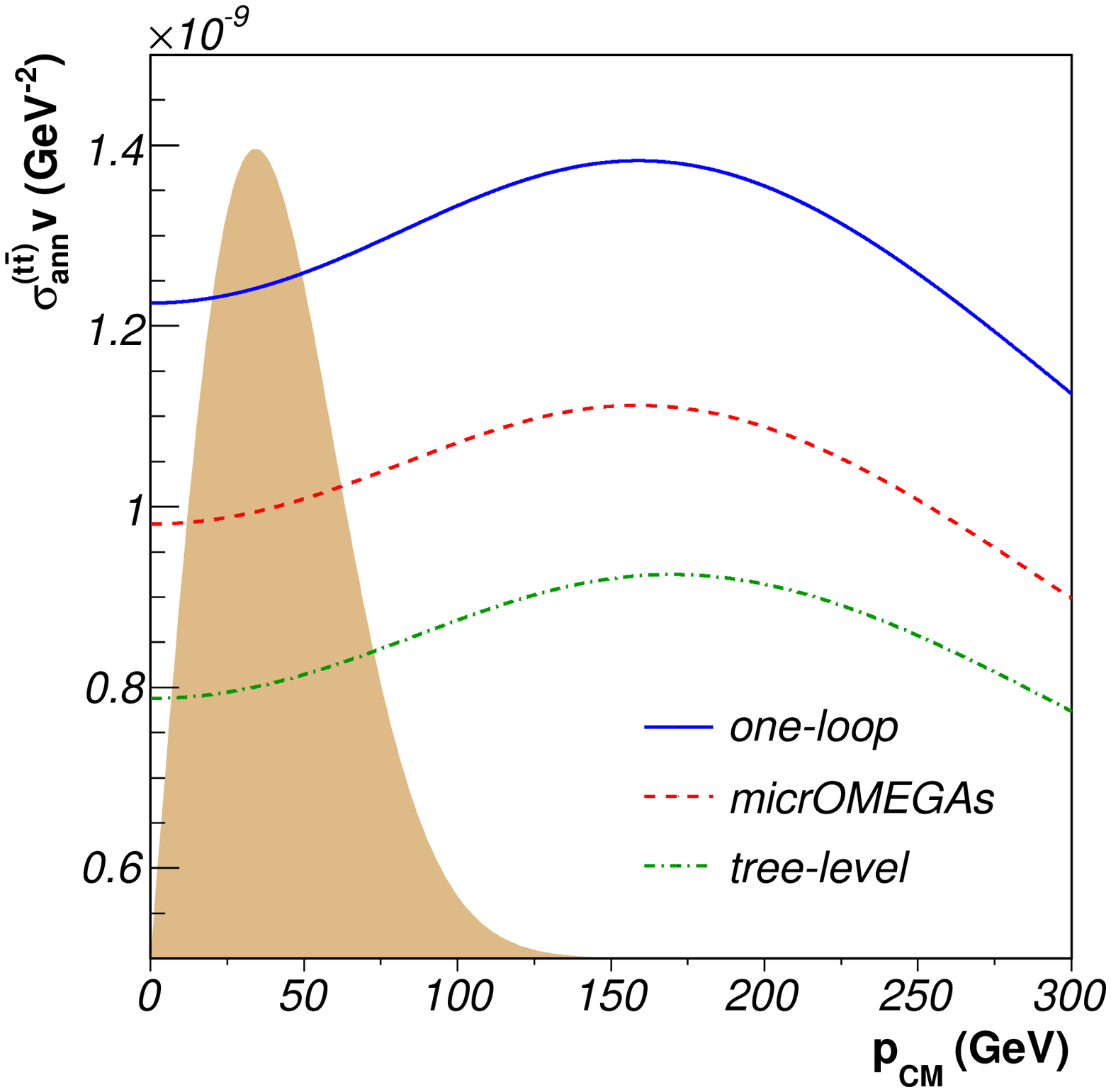}\\
    \includegraphics[scale=0.3]{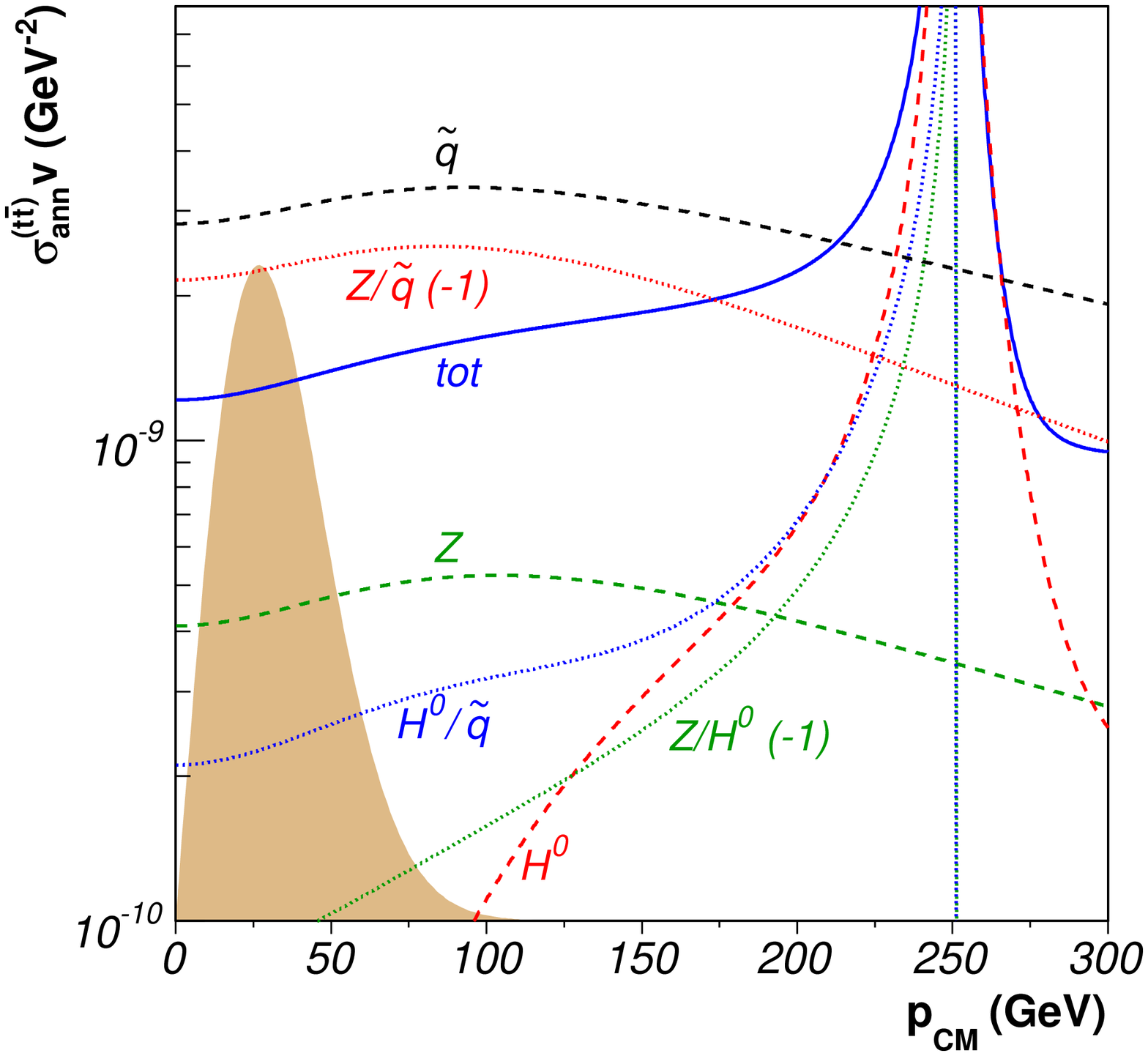} &
    \includegraphics[scale=0.3]{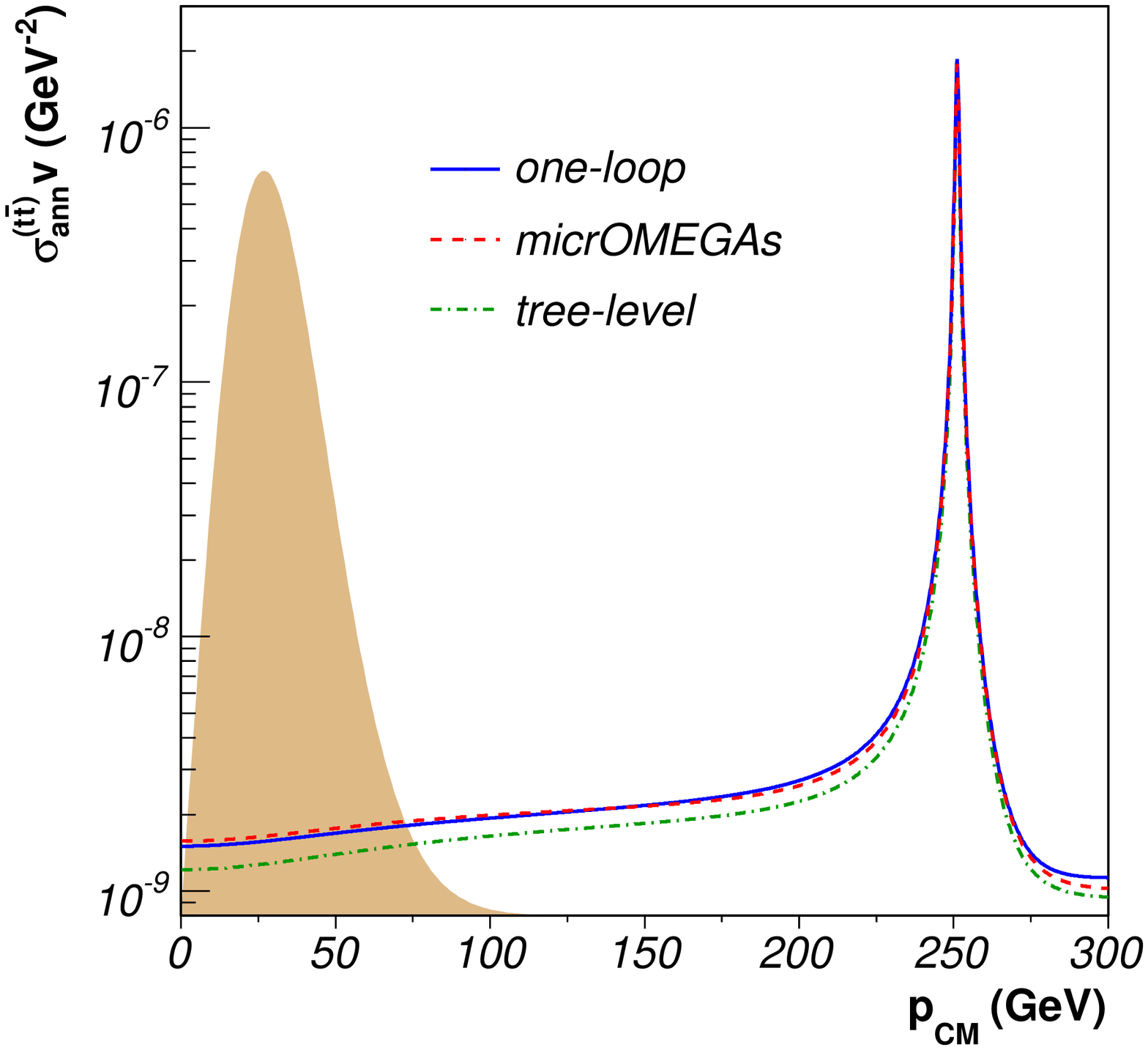}
\end{tabular}
\caption{The contributions of the different diagrams to the annihilation cross section of a neutralino pair into top quark-antiquark pairs (left) and the effect of the radiative corrections on the annihilation cross section (right) as a function of the center-of-momentum energy $p_{\rm cm}$ for the parameter points I (top) and II (bottom). The shaded area indicates the velocity distribution of the neutralino at the freeze-out temperature in arbitrary units.}
  \label{CSfig}
\end{figure}

We investigate two scenarios with non-universal gaugino masses where the relevant parameters are 
given in Tab.~\ref{tab:a}. The common feature of both scenarios is a lower gluino mass $M_3$ which leads 
through the renormalization group evolution and electroweak symmetry breaking to an increased higgsino 
fraction of the neutralino. An increased higgsino fraction enhances the $Z^0$-boson exchange as in our scenario I 
(see top row of Fig.~\ref{CSfig}) and also the squark exchange is larger due to enhanced Yukawa couplings. 
Moreover, the mass of the scalar tops can be lowered by decreasing the scalar mass parameter $m_0$ and using the 
fact that the gluino mass parameter is independent. Very light stops enhance their $t$-channel exchange and lead 
to dominant quark final states as in our scenario II (see bottom row of Fig.~\ref{CSfig}).

\begin{figure}
\begin{tabular}{cc}
  \includegraphics[scale=0.3]{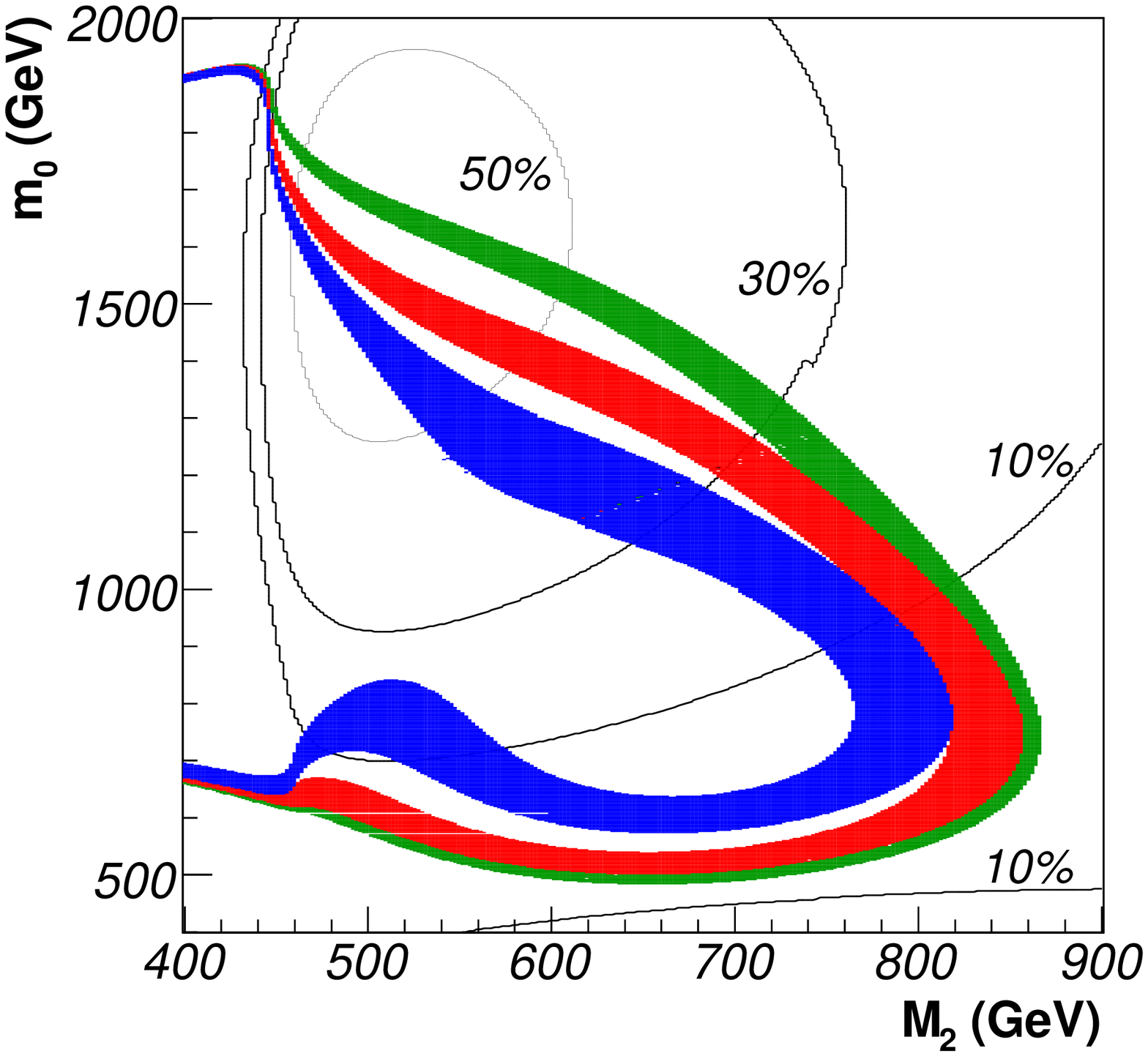} &
    \includegraphics[scale=0.3]{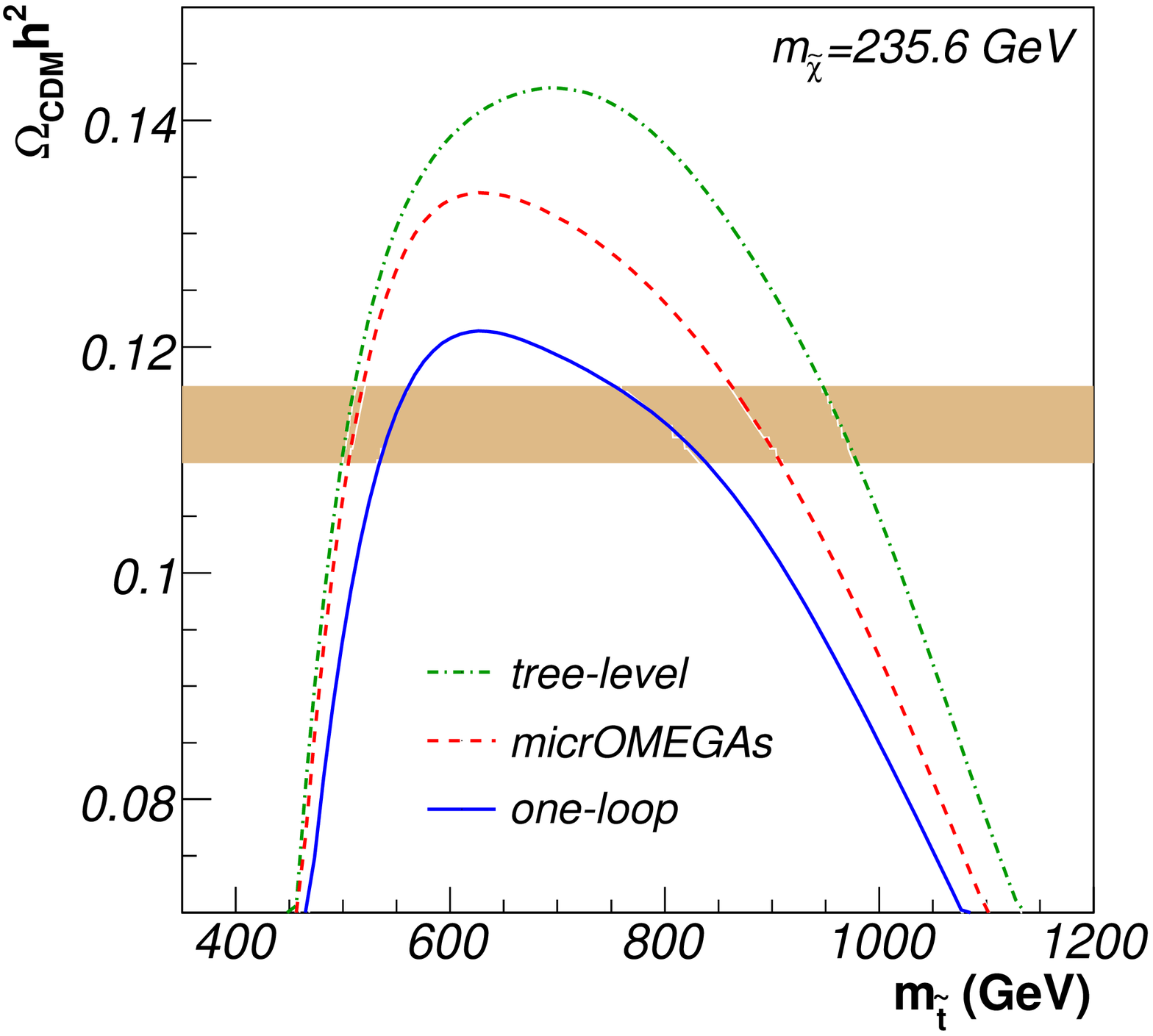}\\
    \includegraphics[scale=0.3]{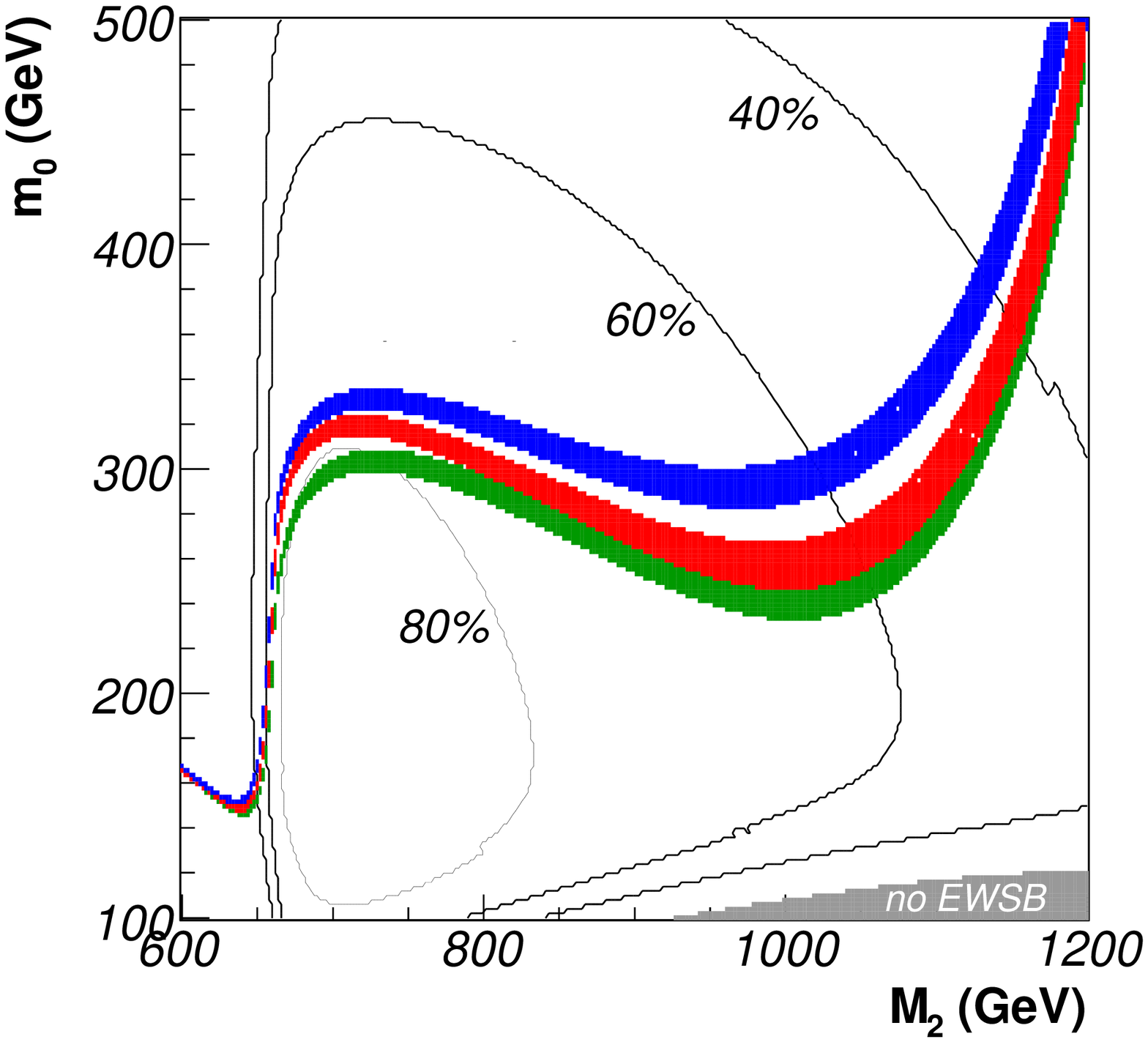} &
    \includegraphics[scale=0.3]{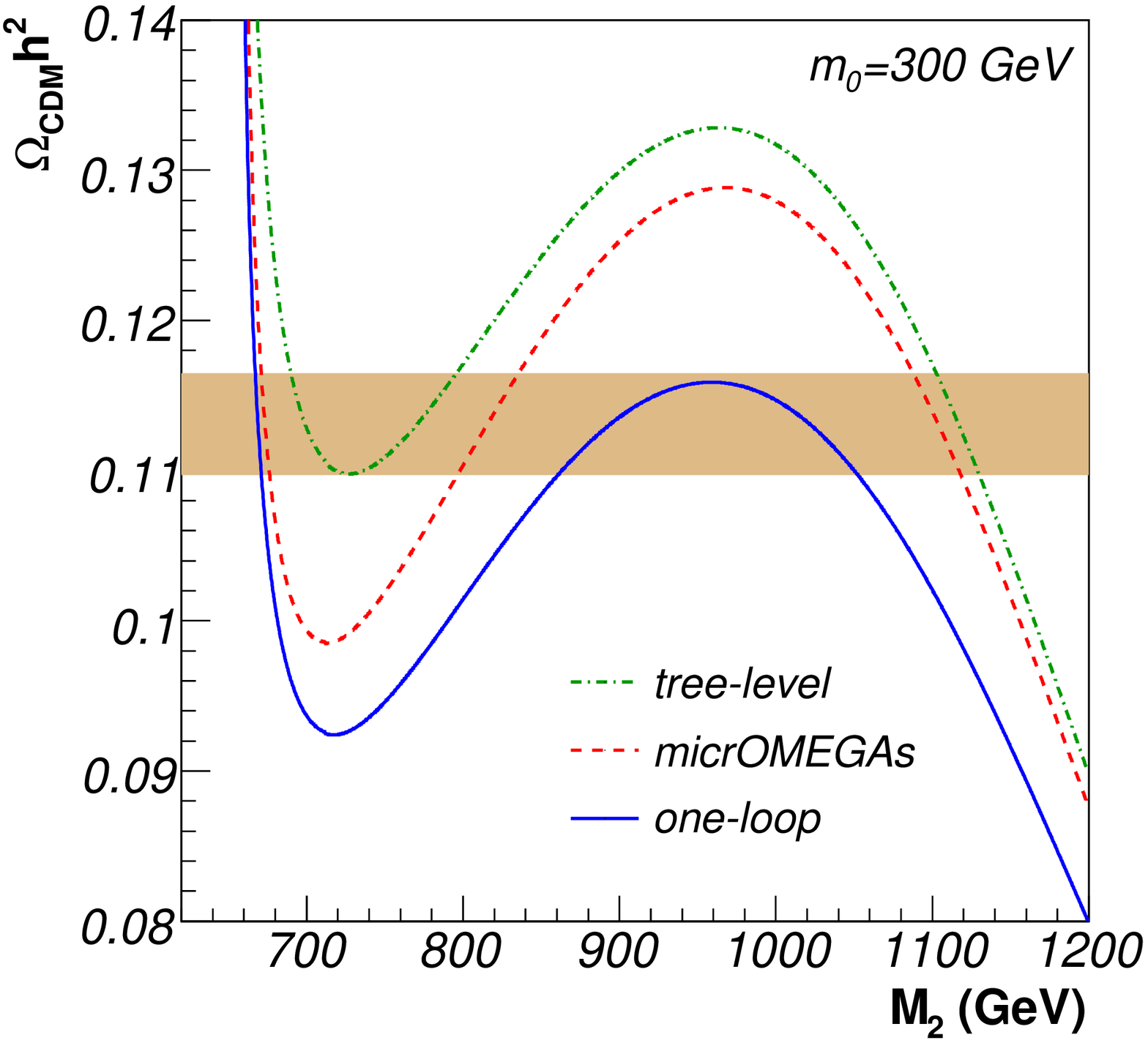}
\end{tabular}
\caption{Cosmologically favored regions in the $m_0$-$M_2$-plane (top and bottom left) for the parameter points I and II. We show the regions that satisfy the constraint from Eq.~(\ref{cWMAP}) for our tree-level calculation (green), the calculation implemented in {\tt micrOMEGAs} (red), and our calculation including the full SUSY-QCD corrections (blue). The prediction of the neutralino relic density $\Omega_{\rm CDM}h^2$ including the tree-level (green dash-dotted) cross section, the approximation included in {\tt micrOMEGAs} (red dashed), and the full one-loop SUSY-QCD corrected cross-section (blue solid) as a function of either the high-scale gaugino mass parameter $M_2$ for fixed $m_0=300$ GeV (top right) or the stop mass $m_{\tilde{t}_1}$ for a fixed neutralino mass $m_{\tilde{\chi}}=235.6$ GeV (bottom right).}
  \label{ScanFig}
\end{figure}

The effects of SUSY-QCD radiative corrections on the relic density (blue) is shown in Fig.~\ref{ScanFig} for both scenarios from Tab.~\ref{tab:a} and compared to the tree-level result (green) and to the approximation implemented in the public computer code {\tt micrOMEGAs 2.1} (red). For both scenarios there are important contributions from squark and $Z^0$-boson exchanges due to the large destructive interference between the two channels. In general, the SUSY-QCD corrections are between 15\%-20\% and are larger than the experimental uncertainty. Therefore, distinct bands appear in scans around our parameter points I and II (see Fig.~\ref{ScanFig}). This demonstrates that the SUSY-QCD corrections are a necessary ingredient when identifying favored regions of parameter space or when extracting the underlying SUSY-breaking parameters from cosmological data. 

\bibliographystyle{aipproc}   % if natbib is available

\bibliography{susy09}

\IfFileExists{\jobname.bbl}{}
 {\typeout{}
  \typeout{******************************************}
  \typeout{** Please run "bibtex \jobname" to optain}
  \typeout{** the bibliography and then re-run LaTeX}
  \typeout{** twice to fix the references!}
  \typeout{******************************************}
  \typeout{}
 }

\end{document}